# Technical Paper Recommendation: A Study in Combining Multiple Information Sources

**Chumki Basu**  CBASU@CS.RUTGERS.EDU
*Department of Computer Science, Rutgers University, 110 Frelinghuysen Road,*
*Piscataway NJ 08854-8019 and*
*Telcordia Technologies, Inc., 445 South Street,*
*Morristown NJ 07960-6438*

**Haym Hirsh**  HIRSH@CS.RUTGERS.EDU
*Department of Computer Science, Rutgers University, 110 Frelinghuysen Road,*
*Piscataway NJ 08854-8019*

**William W. Cohen**  WCOHEN@WHIZBANG.COM
*WhizBang! Labs, WhizBang Labs - East, 4616 Henry Street,*
*Pittsburgh PA 15213*

**Craig Nevill-Manning**  NEVILL@CS.RUTGERS.EDU
*Department of Computer Science, Rutgers University, 110 Frelinghuysen Road,*
*Piscataway NJ 08854-8019*

## Abstract

The growing need to manage and exploit the proliferation of online data sources is opening up new opportunities for bringing people closer to the resources they need. For instance, consider a recommendation service through which researchers can receive daily pointers to journal papers in their fields of interest. We survey some of the known approaches to the problem of technical paper recommendation and ask how they can be extended to deal with multiple information sources. More specifically, we focus on a variant of this problem – recommending conference paper submissions to reviewing committee members – which offers us a testbed to try different approaches. Using WHIRL – an information integration system – we are able to implement different recommendation algorithms derived from information retrieval principles. We also use a novel autonomous procedure for gathering reviewer interest information from the Web. We evaluate our approach and compare it to other methods using preference data provided by members of the AAAI-98 conference reviewing committee along with data about the actual submissions.

## 1. Introduction

We can define the *paper recommendation problem* as follows:

> Given a representation of my interests, find me relevant papers.

In fact, if we replace *papers* in the above definition with the name of some other artifact of choice, we have yet another instantiation of a recommendation problem. What then makes paper recommendation all that interesting?

The ability to automatically filter a large set of papers and find those that are most aligned with one's research interests has its advantages. With the growing number of publications, many of them online, it is difficult to keep up with the latest research, even if it's within one's field. With the timeliness of information becoming all the more critical, it is also desirable for a paper to reach its target audience with minimal latency. Although a straightforward approach to finding relevant papers may look for close matches between a person's interests and a paper's content, what is less clear is how to represent both the interests of the researchers and the contents of the papers.

Another feature that sets paper recommendation apart is that there is a variant problem which must be dealt with on a regular basis by numerous conference chairs. Conferences offer a venue where a large number of fairly specific papers must be distributed to a smaller number of reviewers, all within a very tight timeframe. Even with the scope of the problem being constrained to some degree by topic, conference organizers and/or reviewers still must expend a great deal of time and effort before they can begin the reviewing process. This would suggest that there can be real value in finding ways of automating the filtering process that would make it less burdensome to the potential consumers.

We consider algorithms for recommending focused sets of technical papers. We use conference reviewing as a platform to explore a series of questions relating to the recommendation process. There has been new interest in the AI community for this problem recently since it was proposed as a "challenge" task at IJCAI-97 (Geller, 1997). Our focus on conference reviewing turns out to be a natural choice since we can obtain data about a set of papers, *i.e.*, the conference submissions, and we can also obtain information about the preferences of a set of reviewers for these submissions. In the following section, we discuss related work that addresses the conference reviewing problem. We also consider how other work in the area of recommender systems – *e.g.*, recommending articles to newsgroup readers or recommending Web pages to Web site visitors – can contribute to this task. However, our focus is on varying the sources of information in our data representations, thereby allowing us to formulate different recommendation algorithms based on how we recombine these sources when computing similarity. We show that there is indeed a difference in performance when we vary the amount and source of data, compared to the baseline of using a single source of information in our data representations. We also compare these recommendation algorithms against each other, against collaborative filtering, and against the random assignment of papers to reviewers. We apply our methods to experimental data involving reviewer preferences and conference abstracts for the AAAI-98 conference. [1]

## 2. What We Know about Paper Recommendation

We already know that by recommending papers to reviewers, and more generally, to the arbitrary researcher, we are trying to be selective in choosing those papers that will ultimately reach the consumer based on relevance to interests or expertise. However, finding papers for conference reviewers is necessarily a more complex task, since papers may be assigned to reviewers based on other criteria. For instance, reviewer load balancing and conflict-resolution of reviewer-author affiliations may be two such criteria. In addition, the reviewer's own reviewing preferences may be influenced by considerations such as a paper's readability and novelty. For example, a preference for novelty may lead a reviewer to choose a paper simply because it is not relevant to his or her interests.

---

1. The data were obtained with permission from AAAI, the AAAI reviewers, and when appropriate, from the authors of the submitted papers.

Our methods are not suited to address these latter issues for a number of reasons. First, for confidentiality purposes, we lack information related to the author identity or affiliation of the submitted conference papers. Secondly, since constraint-satisfaction is not our main concern – we are primarily interested in finding the best papers for each person without regard to whether multiple people receive the same paper – we do not incorporate other criteria into our selection procedure. We also do not have a way to represent the "novelty" of a paper with respect to any consumer, and thereby do not have a means for recognizing it. Finally, our methods do not distinguish between the notion of interest and expertise with respect to reviewers. For the more general recommendation problem, the researcher may want to retrieve papers in areas outside of his or her expertise, in which case a separate representation for each would be needed.

Previous work in the area of assigning conference papers to reviewers had approached the problem as one of content-based information retrieval. Dumais and Nielsen (1992) used data provided by 15 members of the reviewing committee for the HYPERTEXT '91 conference. These reviewers not only submitted abstracts of their papers and/or interests, but also provided complete relevance assessments for the 117 papers submitted to the conference. Using an information retrieval method known as latent semantic indexing (LSI), they compared the reviewer abstracts with the submissions, ranking the submissions from most to least similar to each reviewer. From their results, they noticed, based on the performance metric that evaluates the number of relevant articles returned in the Top 10, that they could achieve an average of 48% improvement using their automated methods compared to random assignment of articles to reviewers.

While these results are encouraging, we believe that the widespread availability of online resources introduces opportunities for exploring some new issues. What if the reviewers weren't asked to supply interest information? Can the process of gleaning reviewer interest data be automated with simple methods? How well do we do at retrieving relevant papers using this "approximation" of reviewer interests? The automatic collection of reviewer interest information from the Web, which effectively removes the reviewer from loop, is a novel aspect of our research.

Yarowsky and Florian (1999) attempted a similar task for the ACL'99 conference. However, their primary focus was on classification – the assignment of every paper to exactly one of six conference committees. They used 92 papers which were submitted to the ACL conference in electronic form and also requested committee members to provide representative papers. When the number of papers returned by these members was insufficient, they augmented the collection with other papers downloaded from online sources. They used content-based retrieval (within the context of the vector-space model (Salton, 1989)) as one of their routing strategies. The main algorithm first computed a centroid for each reviewer based on representative papers and then computed a centroid for each committee as the sum of its reviewer centroids. Then, each paper was classified (assigned to a committee) by computing its *cosine similarity* with the committee centroids and choosing the one with the highest rank. Amongst other approaches, they experimented with a *Naive Bayes* classifier and the assessment of similarity between reviewing committee members and authors cited in the papers. Based on their system performance relative to human judges on the same task (evaluated against the actual assignments provided by the program chair of the conference), they extrapolated that automated methods could be as effective as human judges, especially in cases where the judges may be less experienced.

When we are dealing with large conferences with several hundred papers covering a variety of areas, the information load is even greater for conference organizers and reviewers alike. In these cases, getting evaluative relevance judgments for all submitted (or even accepted) papers from the reviewers is not feasible. (As an example, for the AAAI conference, reviewers do not even have to state their preferences for all the papers they can potentially review. Instead, they can stop scanning the list as soon as they have filled up their quota of "bids"– papers they expressed interest in reviewing.) Therefore, we focus on building an extensible framework for recommendation – defining a process whereby we can systematically incorporate more information in formulating recommendation algorithms, for the purpose of generating better recommendations.

Content-based information retrieval, also known as *content-based filtering*, is a popular recommendation method: consider systems that recommend Web pages such as Syskill & Webert (Pazzani & Billsus, 1997). There are a number of other systems such as WebWatcher and Fab that do content-based filtering, mainly as part of a hybrid approach that also involves *collaborative filtering*. Whereas content-based filtering looks only at the *contents* of an artifact (*e.g.*, the words on a Web page), collaborative filtering will also consider the *opinions* of other like-minded people with respect to these artifacts. Collaborative filtering has been used to recommend NetNews articles (Konstan, Miller, Maltz, Herlocker, Gordon, & Riedl, 1997), movies (Hill, Stead, Rosenstein, & Furnas, 1995; Basu, Hirsh, & Cohen, 1998), music (Cohen & Fan, 2000; Shardanand & Maes, 1995), and even jokes (Gupta, Digiovanni, Narita, & Goldberg, 1999). Since both content-based and collaborative methods use data that are orthogonal to one another, there are opportunities to come up with hybrid approaches that use combinations of the data. Our own work on movie recommendation provides another example of how to design a hybrid system. Hybrid systems exploit data from multiple sources with the expectation that they can do better by compensating for the limiting factor of data sparseness associated with any single source.

In our current study, we would like to identify different sources of information to describe both papers and reviewers, with the expectation that the individual pieces themselves, along with knowledge of how to combine them, can make a difference in the recommendations. Although we do share the common goal of combining data from multiple sources with the hybrid recommendation approaches, the algorithms that we develop are strictly content-based. For evaluative purposes, we also compare our algorithms against the results of applying collaborative filtering methods to the set of reviewer preferences.

## 3. Representing Papers and Reviewers

Our approach to recommendation is to represent each entity using a variety of information sources, to enumerate different combinations of these sources, and to evaluate the effectiveness of these combinations using ranked-retrieval methods. For the paper recommendation problem, we have two types of entities — papers and their consumers (reviewers, in our case). For each entity, we can represent the salient features of that entity as a sequence of one or more information sources. In addition, we also need another type of information source that relates a reviewer to a paper, namely, the reviewers' actual preferences for the papers. We begin with a discussion of our choice of information sources — some of these choices are based on data that are typically used to assign papers to reviewers, and are usually provided explicitly by the papers' authors, while other choices rely more on implicit knowledge mined from the semi-structured data available on the Web.

### 3.1 Paper Information Sources

All of our experiments were based on a compilation of submitted abstracts obtained from AAAI for the AAAI-98 conference. There were 466 papers submitted to this conference. AAAI gave us a collection of 256 papers to use in our experiments — the abstracts of 144 accepted papers and the abstracts of 112 papers that had been rejected but whose authors had granted AAAI permission to provide the abstract for this work. Also excluded were any papers that had been authored by any of the authors of this paper.

For each submission we obtained its title, abstract, and a set of user-assigned keywords from a prespecified list. Therefore, each paper has associated with it a set of three information sources all of which were provided by the papers' authors. Although one may consider the body of the paper as another source, this information was not available to the reviewers (nor us), so we do not use it as a source.

### 3.2 Reviewer Information Sources

So far, we have seen an example where an entity such as a paper can be represented by multiple information sources mainly because it is composed of distinct units such as a title, abstract, etc. However, there is another case where we may want to multiply-represent an entity. Consider trying to automatically compose a representation of a reviewer's interests. We may try first to go to the reviewer's home page. From there, we may decide to look around for the reviewer's papers. Each of these sources can offer a different point-of-view of the reviewer's interests, and therefore, can be considered as a separate unit. We focus on these sources – the reviewer's entry-level home page and the papers that are referenced from the home page – as a substitute for asking the reviewer to provide interest information.

We believe both home pages and online papers are credible information sources since it is likely that a fair number of conference reviewers have stated their research interests in either or both sources. Since one of our paper information sources is the paper abstract, we decided to represent the reviewer as an "abstract of interests". In the case of home pages, the entire text of the reviewers's entry-level home page was taken as an abstract of the reviewer's interests. In the case of PostScript files, we define an abstract to be the first 300 words extracted from the paper.

We extracted all of this information from the Web using pre-existing utilities. To find reviewers' home pages, we fed the names and affiliations of the members of the review committee into *Ahoy*,[2] a home page finding engine (Shakes, Langheinrich, & Etzioni, 1997). When *Ahoy* returned at least one match, we supplied the *URL* as a starting point for *w3mir*,[3] an HTTP service that retrieves files from the contents of Web sites. We used

---
2. http://ahoy.cs.washington.edu:6060.
3. http://www.math.uio.no/~janl/w3mir.

*w3mir* to download only HTML files and PostScript files accessible from the entry-level home page and residing on the same site.[4] Since all of a person's papers may not directly be available from one site, we additionally retrieved cross-references to other sites which contained PostScript files, also using *w3mir*. The PostScript files were then converted to ASCII using *PreScript* (Nevill-Manning, Reed, & Witten, 1998).

All PostScript files retrieved for a reviewer are treated uniformly. Although it would be desirable to attempt to do so in future work, we make no attempt to determine the timeliness of a paper, especially with respect to a reviewer's current interests. We also do not distinguish between journal papers, conference papers, and even lecture notes. It is for this reason that we do not attempt to do any detailed analysis of the contents of these files (*e.g.*, to automatically extract titles, abstracts, etc.). Instead, we rely on heuristics such as looking at the first $N$ words to approximate a paper's abstract. Although detailed analysis is likely to be valuable in the paper recommendation process, our immediate goal is to obtain a gross sense of the usability of various sources of semi-structured information.

### 3.3 Reviewer Preferences

To evaluate our queries we need some "ground truth" — some set of data specifying what papers each reviewer had selected as suitable for him or her to review. With this information, we can evaluate how different approaches perform in making the same choices. We note that this is only an approximation to the full set of abstracts that the reviewer might have liked — the reviewing process only requires a reviewer to find some minimum quota of papers, and once that quota is reached, a reviewer need not look at other papers to find more. We view this optimistically as yielding a close approximation to what a reviewer's full set of preferences would be, since reviewers are able to peruse abstracts by keywords and often attempt to inspect at least the subset of papers labeled by keywords in the areas in which they are knowledgeable.

In our experiments ground truth comes from the actual preferences stated by 122 (of the 230) AAAI-98 reviewers who gave AAAI permission to release their preference information for the papers we considered in this work. We point out that this data only reflects the reviewers' initial preferences for reviewing. We do not have data on what papers the reviewers actually received following the AAAI reviewer assignment process.

Of course, one potential limitation of this data is that it is based only on a portion of data that may not be representative of the entire data for the conference. For example, we have preference data for approximately half of all of the reviewers and are predicting preferences for a collection of papers whose distribution is skewed towards the accepted papers. There is also the issue of whether AAAI researchers are representative of the much larger community of researchers at large. (We can ask a similar question of the user populations of other conferences as well). However, we consider these as acceptable limitations resulting from our use of conference reviewing as a platform for paper recommendation.

---

4. At the moment, we focus on PostScript for convenience, but there is no reason to limit ourselves to just one file format; the main constraint is being able to extract the words from a document.

## 4. Recommendation Methodology

In this section, we examine both collaborative and content-based methods of recommendation. These methods allow us to explore the use of different subsets of the data described in the previous section.

### 4.1 Recommending with Reviewer and Paper Information Sources

In the following sections, we outline a content-based recommendation framework that uses data describing the papers as well as data describing the reviewers to make recommendations. The reviewer preference data is then used for evaluation purposes, but not as input to the recommendation process.

In order to locate papers that closely match reviewer interest data we rely on *ad hoc* similarity metrics commonly used in the information retrieval community. We will describe these methods further in the section on WHIRL. In brief, for each reviewer we compare the given reviewer representation with the appropriate paper information source(s). Each of these comparisons can be implemented as a query that returns a rank-ordered list of papers. We can consequently compute *precision* at Top $N$, or the proportion of the papers returned that were actually preferred by the reviewer, for each query. Our final score for each query is the average of this value, computed over a subset of 50 reviewers (from the larger set of reviewers who gave us their permission).

Our recommendation algorithms take different paper and reviewer information sources as inputs. Since our data can be plotted along two dimensions, let *Reviewer* be the set of information sources describing reviewers and *Paper* the set of information sources describing papers. We can construct a *Reviewer* × *Paper* matrix where each entry in this matrix is a score measuring the effectiveness of using the respective sources, ($Reviewer_i$, $Paper_j$), to compute similarity between reviewers and papers when performing a ranked-retrieval. For instance, given the paper and reviewer representations we have described, we can construct a 2 × 3 matrix, which gives us 6 possible evaluations or scores. We will refer to this matrix as the *recommendation sources matrix*.

Conceptually, we can extend the recommendation sources matrix along each dimension, by considering combinations of the rows and columns. We refer to the augmented matrix as the *source combinations matrix*. We can now define a recommendation algorithm as a combination method or procedure applied to one or more rows/columns of the source combinations matrix. This introduces another dimension for comparison – the *combination method* itself – which we consider by looking at replicates of the source combinations matrix.

Now, we can pose the following questions for experimental analysis:

- Do recommendation algorithms that incorporate more information lead to better performance?

- If so, does the method of combining data used by the algorithm make a difference?

### 4.1.1 WHIRL

For all of our queries, we use WHIRL, a system specifically designed for information-integration tasks (Cohen, 1998b; Cohen & Hirsh, 1998). For these tasks, it is often necessary to manipulate in a general way information obtained from many *heterogeneous* online sources, each potentially having its own data organization and terminology. In particular, WHIRL makes it possible to integrate information that can be decomposed and represented in a clean, modular way. For example, we would like to have information about home pages and PostScript papers represented separately, using the information integration tool to resolve these sources of information.

WHIRL is a conventional DBMS that has been extended to use *ad hoc* similarity metrics developed in the information retrieval community. Using these metrics, it can reason about pieces of text culled from heterogeneous sources based on the *similarity* of values rather than on strict equality. WHIRL computes similarity using the "vector-space" representation to model text (Salton, 1989). Each text object is represented by a vector of term weights (where the terms have been stemmed using Porter's algorithm (Porter, 1980)) based on the *TFIDF* weighting scheme. Similarity between two vectors is computed using the *cosine similarity* metric. The answers to a query are presented by rank-ordering the generated tuples, with tuples having more similar pairs of attribute fields appearing first.

For example, using WHIRL, we can pose the following query:

>       SELECT Reviewer.Name, Paper.ID
>         FROM Paper AND Reviewer
>        WHERE Reviewer.Descriptor SIM Paper.Abstract

This query will return a list of reviewer names and paper IDs for papers whose abstracts were similar to the reviewer's interest descriptor. Rather than returning only those tuples for which the descriptor and abstract fields are identical, as would be performed by a traditional database join, this query returns Name and ID pairs for those tuples whose fields contain similar terms, ordered according to decreasing value of similarity. The advantage of doing *ad hoc* joins without requiring the textual fields to be identical to one another is important when the text comes from multiple sources and thereby may use different terminology. It is also important from the perspective of comparing the relative importance of different fields to one another in an efficient way.

To use WHIRL all data must be stored in the form of WHIRL relations. For our data we constructed two relations, each one representing different information sources. For each conference submission, we form a Paper relation containing its id, abstract, keywords, and title. For every reviewer, we form a Reviewer relation which contains a single tuple with attributes representing the reviewer's name and some representation of the reviewer's interests (for example, based on the reviewer's home page).

So far, we have discussed how we can use WHIRL to formulate queries involving a single information source for both reviewers and papers. However, an advantage of the WHIRL approach lies in the simplicity with which we can extend these queries to incorporate multiple sources. The primary advantage of using WHIRL in our work is the ease with which we can measure the impact of *conjunctive queries* incorporating data from multiple sources. We form conjunctive queries by adding multiple conditions to a WHERE clause:

>       SELECT Reviewer.Name, Paper.ID
>         FROM Paper AND Reviewer
>        WHERE Reviewer.Descriptor SIM Paper.Abstract
>          AND Reviewer.Descriptor SIM Paper.Keywords

When a WHIRL WHERE clause contains multiple conditions, the similarity scores of the individual conjuncts are combined by taking their product as though they were independent probabilities. Since similarity scores are not independent probabilities, we only use it as a convenient way to combine scores, albeit one that offers a straightforward approach to combination which has been previously studied (Cohen, 1998a). In the above query, WHIRL would assign a score that reflects both the similarity of the submitted paper's abstract and the reviewer's descriptor, as well as the similarity of the submitted paper's keywords and the reviewer's descriptor.

### 4.1.2 Combining Information Sources by Query Expansion

What does it mean for a recommendation algorithm to combine data from multiple information sources? This means enumerating the information sources that can be used as possible inputs to the algorithm, and then defining a way to use these sources to compute similarity. For instance, suppose we look at 1 reviewer source and 2 paper sources for a given collection of reviewers and papers. To decide whether a paper is likely to interest the reviewer, we can compute the similarity between the reviewer source and each of the paper sources and combine the two similarity scores. Alternatively, we can compute a single similarity score by first combining the two paper sources into a single representation and then computing its similarity with respect to the reviewer source.

The idea of combining two sources into a single representation can be implemeted by appending terms from the sources. In information retrieval, terms from relevant sources are often appended to a baseline representation of a query during the process of query reformulation. This is usually referred to as *query expansion*. Since our methods bear a resemblance to query expansion, we make this analogy. These expansion methods will be further described in the following sections. Of course, we do not have prior knowledge of the relevance of our sources, and in this sense, we differ from the information retrieval implementation of query expansion.

When we compare the relative performance of recommendation algorithms, we have multiple dimensions along which to compare the results. We can differentiate the results based on the methods used to combine the data and compute similarity or we can differentiate between the results based on which information sources were used in the comparison. In other words, on the same set of inputs, does one method of query expansion perform better than another? If we want to compare the merit of a single source, we can consider two groups of algorithms – those that include a given source as input to the algorithm, and those that exclude this source. If we simply count the number of times algorithms that include this source outperform algorithms that exclude it, we can determine the relative merit of the source.

### 4.1.3 The Concatenation Method

One way to "add" information from a new data source is to append the terms appearing in the source to the original WHIRL query. For this type of query, we always have a single WHIRL conjunct but each of the textual fields appearing in the conjunct can "grow" with the addition of new terms. We call this method, *queryConcat*.

Suppose, for example, that we start with the base query from the previous section that only compares reviewer descriptors with paper abstracts. Now, suppose we want to compare reviewer descriptors not only to the paper abstracts but also to the paper keywords. One way to do this is to use the *queryConcat* method. We form a new field representing the union of the words appearing in the paper abstract and paper keywords fields which we can substitute in the original query. Let *Paper.Descriptor* = *Paper.Abstract* ∪ *Paper.Keywords*. Our new query is:

> SELECT Reviewer.Name, Paper.ID
>   FROM Paper AND Reviewer
>  WHERE Reviewer.Descriptor SIM Paper.Descriptor

Similarly, we can replace *Paper.Descriptor* in the WHERE clause to represent different combinations of the fields, *Paper.Abstract*, *Paper.Keywords* and *Paper.Title* using the union operator.

### 4.1.4 THE CONJUNCTION METHOD

As we previously stated, an important motivation for using WHIRL is its ability to execute *conjunctive* queries, which we can also use to combine information sources in the recommendation process. For this type of query, instead of adding terms to any particular text field, we add conjuncts to the original WHERE. We refer this method of reformulating queries as *queryConjunct*.

We enumerate the query combinations that we considered for *queryConjunct* as follows. Using the same sources as for *queryConcat*, we can begin the queries as before,

> SELECT Reviewer.Name, Paper.ID
>   FROM Paper AND Reviewer
>  WHERE

but now, replacing the body of the *WHERE* clause with the following:

> A: Reviewer.Descriptor SIM Paper.Abstract
>
> K: Reviewer.Descriptor SIM Paper.Keywords
>
> T: Reviewer.Descriptor SIM Paper.Title
>
> AK: Reviewer.Descriptor SIM Paper.Abstract
>   AND Reviewer.Descriptor SIM Paper.Keywords
>
> AT: Reviewer.Descriptor SIM Paper.Abstract
>   AND Reviewer.Descriptor SIM Paper.Title
>
> KT: Reviewer.Descriptor SIM Paper.Keywords
>   AND Reviewer.Descriptor SIM Paper.Title
>
> AKT: Reviewer.Descriptor SIM Paper.Abstract
>    AND Reviewer.Descriptor SIM Paper.Keywords
>    AND Reviewer.Descriptor SIM Paper.Title

We assign the labels, A (abstract), K (keywords), and T (title) to the queries to identify the paper sources used. (We use these labels in a comparable fashion for the *queryConcat* method, representing the information sources that are concatenated together.)

For each of the above queries, we can also vary the source of data used to represent the reviewers. The first variant accounts for the case where the reviewer's descriptor contains the words from the reviewer's home page; the second accounts for the case where the descriptor contains the union of the first 300 words extracted from each PostScript file obtained from the reviewer's Web pages.

We decided to try yet another combination to see whether using both representations for reviewers would improve performance. For simplicity, we chose to test this hypothesis with an expanded conjunctive query involving a single extra conjunct. We constructed a Reviewer table that contains two attributes: Papers (consisting of the abstracts of the reviewer's PostScript papers) and Homepage (consisting of the reviewer's home page). We then ran each of the above queries, but now with an additional conjunct appearing in each WHERE clause:

> Reviewer.Homepage SIM Paper.Keywords

We chose to use *Keywords* as the *Paper.Descriptor* based on our intuitions that a paper's keywords and a reviewer's homepage would have a greater number of words in common.

### 4.2 Recommending with Reviewer Preferences

Since we have evaluations from the reviewers on a common set of papers, one approach for recommending papers would be to take this information and use it for collaborative filtering. We note that for the actual conference reviewing problem, collaborative filtering as a method for assigning papers may not be practical. Although we have the benefit of using all of the preferences for a set of reviewers in our study, this information will generally not be available to the reviewers as they are making their selections, thereby making it more difficult to base predictions on the preferences of others. Nevertheless, it is worthwhile to measure the impact of using reviewer preferences for the purpose of recommending papers.

The recommendation methodology for the collaborative filtering approaches is implemented as follows: each reviewer is presented with a recommended paper in an online manner. After the paper is presented the reviewer tells the system if the paper was relevant. If it was, then the paper is assigned a rating of 1 and the paper is said to be rated positively. If the paper was not relevant, it is assigned a rating of 0 and is said to be rated negatively. Let $Rating(R, P)$ represent the rating that has been assigned to paper $P$ by reviewer $R$. When the paper is not relevant, the reviewer also provides a single relevant paper as a positive example in order to condition future recommendations. Since we know which papers were liked by the reviewers, we can simulate this process with the data that we have. We experiment with two collaborative filtering algorithms: *kNN* (Hill et al., 1995; Cohen & Fan, 2000) and *Extended Direct Bayes* (Cohen & Fan, 2000). We let $P_1, P_2, ..., P_{t-1}$ represent the papers that have been previously rated by the reviewer in $t-1$ trials. The *kNN* algorithm uses the following distance metric to locate other reviewers, $R_i$, closest to the current reviewer with respect to the papers that have already been rated:

$$Dist(R, R') = |Rating(R, P_1) - Rating(R', P_1)| + ... + |Rating(R, P_{t-1}) - Rating(R', P_{t-1})|$$

We can then compute a score for an arbitrary paper, $P$, with respect to the ratings of the $k$ closest reviewers, $R_1, ..., R_k$, as follows:

$$Score(P) = Rating(R_1, P) + ... + Rating(R_k, P)$$

According to the above methodology, the highest scoring paper will be presented to the reviewer as the next recommendation.

*Extended Direct Bayes* can be viewed as an *ad hoc* extension of a direct Bayesian approach to recommendation. We define $R(P_i, P_j)$ to represent the Laplace-corrected estimate of the prior probability that the reviewer will give $P_j$ a positive rating. ($R(P_i, P_j)$ can be thought of as measuring the "relatedness" between two papers.) Now consider an arbitrary trial $t$ and let $P_1, P_2, ..., P_{t-1}$ represent the papers that have been rated *positively* by the reviewer on previous trials and consider an arbitrary trial $t$.

We can now use the following scoring function to rank each paper $P$:

$$Score(P) = 1 - ((1 - R(P, P_1)) \times ... \times (1 - R(P, P_{t-1})))$$

The subtrahend in the above expression represents the probability that $P$ is not related to any $P_i$ (assuming that the $P_i$'s are independent).

### 4.3 Evaluation Methodology

In the following sections, we evaluate the performance of our recommendation algorithms. For collaborative filtering, we compute recommendations for a reviewer until we run out of positive examples to use as feedback. For each reviewer's list of recommendations, we measure precision in the Top $N$; this gives us the proportion of the items returned in the Top $N$ for a given reviewer that were actually preferred by the reviewer. Although it is possible to use other evaluation metrics, we compute precision at different levels of papers returned since it is well-suited to the conference reviewing task. Since a reviewer may get a list of about 10 papers to review, we would like to simulate this by recommending the Top 10 papers returned by our methods. By computing precision, we measure the percentage of papers in this list that would have matched the reviewer's preferences. This metric is also commonly used in the literature. For instance, Dumais and Nielsen (1992) mostly used this measure, *i.e.*, the number of relevant articles in the Top 10, when reporting their results since this constituted a reasonable reviewer load. We additionally report results of precision at Top 30. For the *kNN* algorithm, we set $k = 10$ for our experiments.

Our recommendation algorithms can be seen as a choice of a query expansion method crossed against a choice of the input data sources. For each of the methods *queryConjunct* and *queryConcat*, we ran $3 \times 7$ queries detailed in the previous section. This resulted in 21 runs per reviewer, per method. Each run returned an ordered list of paper IDs. For each run, we again measure precision in the Top $N$ (for $N = 10$ and $N = 30$). In our discussion, we refer to a run using abstracts based on a reviewer's papers as a $p$ run. Similarly, $h$ runs will be based on a reviewer's home page. Finally, $ph$ runs combine both sources of information (using the extra conjunct). The results we will report represent precision values averaged across the reviewers. In order for us to compare performance *across* different information sources, we need to do our evaluation using the same population of reviewers. Not all of the reviewers who provided preference data had home pages and/or papers available online. Therefore, we performed a set of runs using 50 reviewers randomly chosen from the set of

TECHNICAL PAPER RECOMMENDATION: A STUDY IN COMBINING MULTIPLE INFORMATION SOURCES

| Source(s) | A | K | T | AK | AT | KT | AKT |
|---|---|---|---|---|---|---|---|
| p(Top10)  | 0.248 | 0.260 | 0.234 | 0.266 | 0.274 | 0.308 | 0.330 |
| h(Top10)  | 0.210 | 0.284 | 0.232 | 0.288 | 0.270 | 0.320 | 0.332 |
| ph(Top10) | 0.334 | 0.304 | 0.332 | 0.312 | 0.342 | 0.286 | 0.374 |
| p(Top30)  | 0.194 | 0.201 | 0.177 | 0.198 | 0.195 | 0.220 | 0.232 |
| h(Top30)  | 0.169 | 0.217 | 0.183 | 0.226 | 0.199 | 0.232 | 0.232 |
| ph(Top30) | 0.245 | 0.219 | 0.233 | 0.224 | 0.241 | 0.211 | 0.249 |

Table 1: Average Precision Scores at Top 10 and Top 30 Papers Returned using *queryConjunct*.

reviewers who had both home pages and papers available online, and report results averaged across these 50 reviewers.

As we mentioned earlier, reviewer choices may be influenced by a variety of factors ranging from a person's curiosity to a paper's readability. Many of these factors are difficult to model. Furthermore, human judges may assign papers to reviewers according to criteria other than relevance of paper contents to reviewer interests, and their individual opinions may vary. Therefore, it is highly unlikely that our proposed methods will achieve 100% precision. Unfortunately, given the nature of the problem, we have not been able to get an assessment of how human judges would have done at the same task. Nevertheless, we can evaluate our recommendation framework built on content-based information retrieval principles and compare relative performance to other reasonable baseline approaches.

## 5. Results

There are a number of questions we would like to keep in mind as we analyze the results. In the course of our experiments we vary both the amount of information input to our algorithms and the method of query expansion used by the algorithms. One of the questions we would like to answer is what algorithm or set of algorithms is most suited to the task at hand? We also ask whether the choice of inputs results in measurable differences in performance. The tabulation of results which provides the basis for analyzing our content-based algorithms is presented in Table 1 and Table 2. The baseline method against which we compare all algorithms is *random assignment*. This method assigns each reviewer a random collection of papers. With this method, we can expect a precision of 7.0%. In other words, this means that if we were to select papers randomly, on average, each reviewer would like fewer than 1 out of 14 of the papers selected.

Table 1 and Table 2 are replicates of the source combinations matrix we had discussed earlier. Since we ran two trials for Top $N$ papers returned, each table is actually the concatenated representation of the matrices for the Top 10 and Top 30 experiments. In the first three rows of Table 1 and Table 2, we report precision figures of the Top 10 papers returned for the *queryConjunct* method and the *queryConcat* method, respectively.

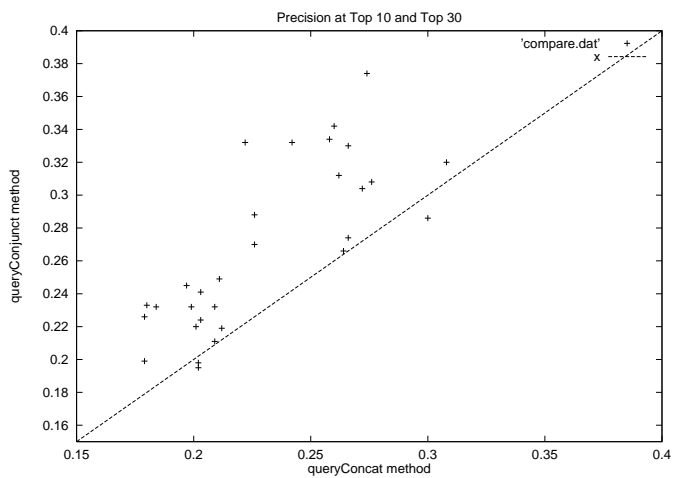

Figure 1: A Comparison of Two Query Methods

Similarly, we show the results for Top 30 papers returned in the bottom three rows of the tables. Since we can view the rows as representing the reviewer sources used in a query and the columns representing the paper sources, we can measure the impact of adding data in two ways. By reading across a row, across groups of columns representing $N$ information sources, we can gauge how the results vary as more paper data are included in the queries. Similarly, by reading down a column, we can gauge the differences in the results as more reviewer data are included in the queries.

Given this information, what can we say about the performance of our recommendation algorithms that used different methods of query expansion? We can compare the relative performance of the two methods *queryConjunct* and *queryConcat* based on the values listed in Table 1 and Table 2. Note that in all cases performance of these methods exceeds that of random selection, with accuracies a factor 2 to 5 times better. In Figure 1, we record this information as a data point for every query that uses two or more sources of information (since the methods differ in how they combine data from two or more sources, it is meaningless to plot points that refer to queries using a single source). In this figure, the x-axis represents queries expanded using the *queryConcat* method and the y-axis represents queries expanded using the *queryConjunct* method. If a point falls on the $x = y$ line, then the two methods yielded the same performance for a query using the same information sources. All points that fall in the area above the $x = y$ line mark those queries where *queryConjunct* had higher precision than *queryConcat*. The data reveal that in almost all cases, *queryConjunct* had higher precision than *queryConcat*, thereby making *queryConjunct* the dominant of the two query expansion methods and the preferred method of the two for the task at hand.

Our expectation is that as we increase the source data we should notice an increase in precision. Specifically, we note that for *queryConjunct*, the query that uses the most information for a paper submission in a majority of cases performs statistically significantly

TECHNICAL PAPER RECOMMENDATION: A STUDY IN COMBINING MULTIPLE INFORMATION SOURCES

| Source(s) | A | K | T | AK | AT | KT | AKT |
|---|---|---|---|---|---|---|---|
| p(Top10)  | 0.248 | 0.260 | 0.234 | 0.264 | 0.266 | 0.276 | 0.266 |
| h(Top10)  | 0.210 | 0.284 | 0.232 | 0.226 | 0.226 | 0.308 | 0.222 |
| ph(Top10) | 0.258 | 0.272 | 0.242 | 0.262 | 0.260 | 0.300 | 0.274 |
| p(Top30)  | 0.194 | 0.201 | 0.177 | 0.202 | 0.202 | 0.201 | 0.209 |
| h(Top30)  | 0.169 | 0.217 | 0.183 | 0.179 | 0.179 | 0.199 | 0.184 |
| ph(Top30) | 0.197 | 0.212 | 0.180 | 0.203 | 0.203 | 0.209 | 0.211 |

Table 2: Average Precision Scores at Top 10 and Top 30 Papers Returned using *queryConcat*.

better[5] than queries that use less information and in no case performs statistically significantly worse.

We should note that adding information will not always lead to monotonically better results. Notice that for *queryConjunct*, in the case of Top 30 papers returned, *hKT* is indistinguishable from *hAKT*. We also note that *phT* performs better (though not statistically significantly better) than *phKT*. There are similar cases for *queryConcat*. How do we explain these gaps? If these are indeed gaps, *i.e.*, they are true statistical differences, then we may consider as an explanation that adding information may also be increasing the amount of noise in our representations. Consider, for example, that keywords from a fixed list can often be a poor match to the real subject matter of a paper. In these special cases, the use of keywords as a source could lead to a degradation in retrieval performance.

Analogous to our analysis of the paper sources, we can now examine any column of Table 1 or Table 2 and measure the effect of adding more information to the reviewer representation. For *queryConjunct*, a majority of the time, we find that queries incorporating more information (*ph* entries) perform statistically significantly better than single source queries (*p* and *h* entries).

So far, we have illustrated how we can move across groups of columns or blocks of rows in the source combinations matrix, adding sources to the queries until there is no improvement. How significant are the gains that we can realize when we do this? Focusing on *queryConjunct*, for every reviewer source, we consider queries that contained data from a single paper source and had the lowest precision. We pair each of these queries with the corresponding query in the same row of the matrix that made use of all of the paper sources and report the resulting improvement in precision in Table 3. For the Top 10 results, we note that in the best case, we can gain an improvement in precision of 58% when going from a single-source to a multi-source query, and for the Top 30 results, we gain an improvement of 37%. These results do support our intuitions that by incorporating more information in our queries, the quality of the retrieval results improves. Since we have a different paper source for the single-source queries in each row of Table 3, we also note that the impact of any given paper source is dependent on the reviewer representation that we use.

---

5. All comparisons between two queries $Q_i$ and $Q_j$ were made using a two-tailed sign test. Specifically, we consider the set $R_{ij}$ of reviewers $r$ for which $precision(Q_i, r) \neq precision(Q_j, r)$ and then estimate the probability
$$p_{ij} = Prob(precision(Q_i, r) > precision(Q_j, r) \mid r \in R_{ij})$$
We consider a difference to be statistically significant if one can reject with confidence $> 0.95$ the null hypothesis that $p_{ij}$ was generated by $\mid R_{ij} \mid$ independent flips of a fair coin.

| Single-Source Queries | Improvement After Adding Two Sources |
|---|---:|
| pT(Top 10) | 41% |
| hA(Top 10) | 58% |
| phK(Top 10) | 23% |
| pT(Top 30) | 31% |
| hA(Top 30) | 37% |
| phK(Top 30) | 14% |

Table 3: A Comparision of Single-Source vs. Multi-Source Queries.

| $Methods(s)$ | $Top\ 10$ | $Top\ 30$ |
|---|---|---|
| $kNN$ | 0.294 | 0.154 |
| $ExtendedDirectBayes$ | 0.300 | 0.129 |

Table 4: Average Precision Scores at Top 10 and Top 30 Papers Returned using Collaborative Filtering Methods.

Can we still come up with an assessment of which sources are significant for the conference reviewing task? For *queryConjunct*, we present a series of figures (Figure 2 to Figure 6) that illustrate the impact of each source by plotting precision values of queries that exclude the source along the x-axis and precision values of queries that include the source along the y-axis (for both $N = 10$ and $N = 30$). If a point falls on the $x = y$ line, then the queries have exactly the same performance — the choice of source is irrelevant. All points that fall in the area above the $x = y$ line mark those queries that had higher precision compared to their query counterparts which did not contain the source.

By simply counting the number of times the queries that include a source outperform the queries that did not include the source, we have one way of ranking the sources in decreasing order of importance. In this case, queries that include the abstract source for papers and the home page source for reviewers have the highest rates of success (when compared to the other information sources for papers and reviewers, respectively).

Now, the natural question to ask is whether the trends that we noticed for *queryConjunct* also hold for *queryConcat*. The answer is no, which also means that *queryConcat* does not give us a definitive answer to the question of whether more information is really better. Just as we have noticed that query performance is linked to both the reviewer and paper sources, we also find that it is linked to the query expansion method.

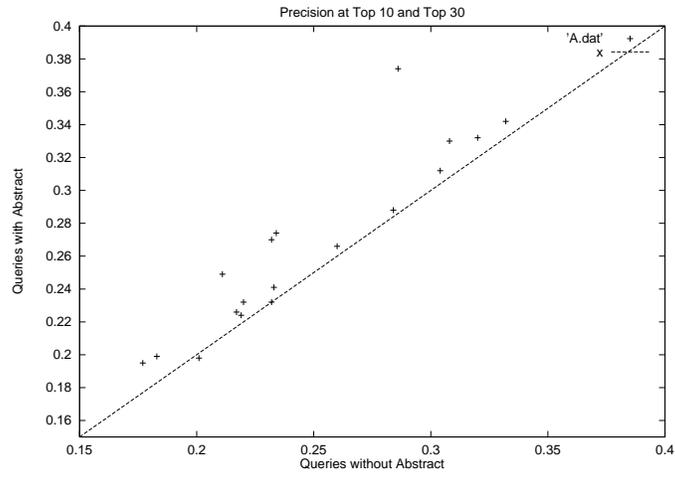

Figure 2: The Role of Abstract as an Information Source

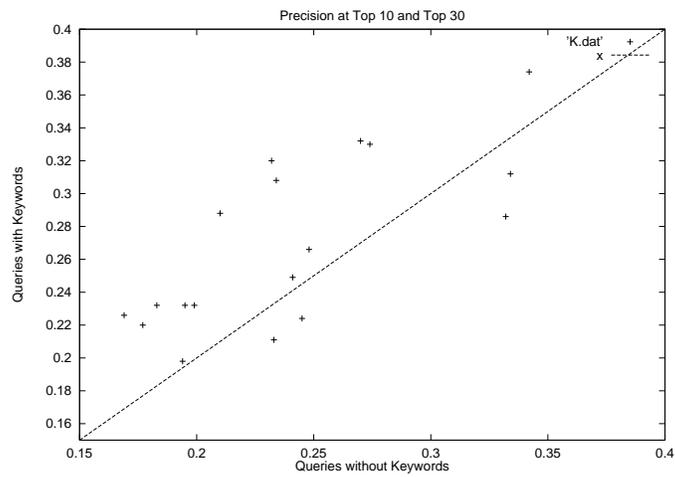

Figure 3: The Role of Keywords as an Information Source

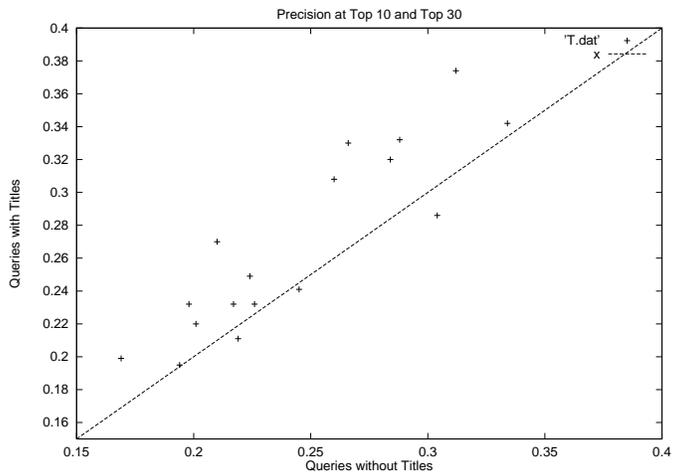

Figure 4: The Role of Title as an Information Source

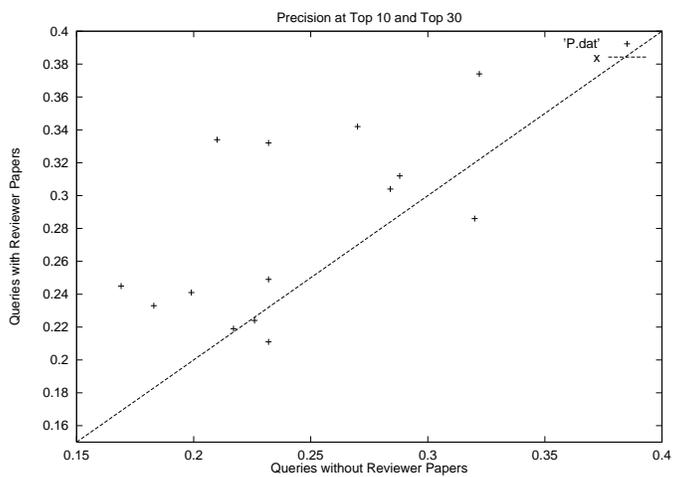

Figure 5: The Role of Papers as an Information Source

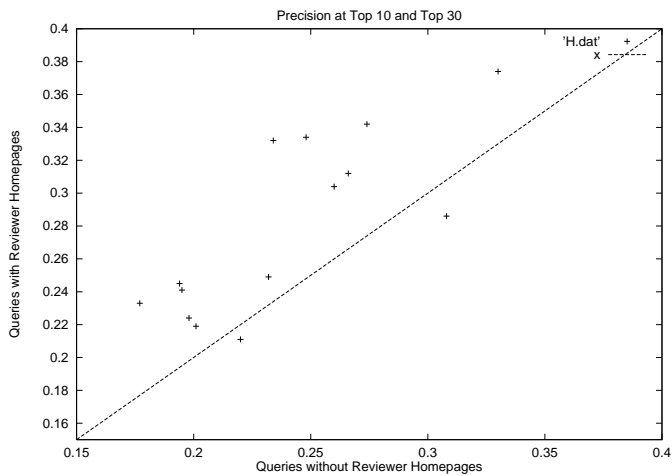

Figure 6: The Role of Homepage as an Information Source

In Table 4, we show the results of the collaborative filtering runs. We report averages of the precision values computed for the Top $N$ (for $N=10$ and $N=30$) papers returned based on the reviewer recommendation lists. Since we stop recommending after we have exhausted the set of positive examples for a reviewer, the reviewer recommendation lists are of varying lengths. In those cases where the size of the list is less than $N$, we still compute precision at Top $N$, assuming the remaining items are incorrect predictions. Both methods for collaborative filtering exceed random selection by a significant margin.

For Top 10 papers returned, the collaborative recommendation methods are competitive with the best performance of *queryConcat*. This is already an interesting observation, since not only do the methods differ, but each method is using different data to make recommendations. We can further state than when we use *queryConjunct* and all information sources to recommend 10 papers, on average almost four papers coincide with the reviewer's preferences. Compared to random selection, collaborative filtering, and *queryConcat*, this method yields more papers of interest to reviewers.

In summary, what have we learned from our experiments? We have found that within the context of peer reviewing of papers, we can make the recommendation process less "people intensive". Most recommendation systems require their users to provide samples of their preferences which are then used to extrapolate future behaviors. Collaborative methods go even further by using preference information across multiple users to predict the preferences of a single user. By automatically collecting reviewer interest information from Web sources and precomputing similarities between these profiles and paper content, we require less input from the reviewers. Furthermore, our content-based retrieval methods can exceed the performance of collaborative methods in this task.

We also believe that our recommendation framework provides an extensible way of formulating queries that provides more control over the *information content* of the queries. We can control not only how much information we include in our queries but also how we incorporate that information. As new data become available, we can evaluate which data sources and/or combinations are more effective, thereby fine-tuning the query formulation process.

## 6. Related Work on Query Reformulation

Since our work on expanding queries using WHIRL can be viewed as a type of query reformulation, we review some related work in the information retrieval community on this topic. Salton (1989) describes the process of query reformulation as that of "moving" a given query towards the relevant items and away from the nonrelevant ones. In the context of the vector-space model of retrieval, this means that given a query expression of the form (Salton, 1997):

$$Q_0 = (q_1, q_2, ..., q_t)$$

where $q_i$ is a number between 0 and 1 representing the weight assigned to $term_i$, we want to arrive at a new query expression:

$$Q'_0 = (q'_1, q'_2, ..., q'_t)$$

such that the weights are adjusted so that new terms can be introduced into the vector representation, while other terms can effectively be removed by reducing their respective weights to 0.

Harman (1992) describes the operational procedure underlying this process as the merging of document and query vectors. More specifically, this means that query terms not in the original query but appearing in the relevant documents are added to the initial query expression. The expansion occurs using both positive and negative weights, depending on whether the terms appears in a relevant or non-relevant document.

The above description assumes that we have relevance judgments for documents that the system can return. Practically speaking, this type of information is hard to come by. Therefore, people have been seeking to compensate for this lack of information by expanding queries using a variety of techniques such as the use of thesauri and relevance feedback. In the latter case, query reformulation is part of an iterative and interactive process whereby users are presented with the results of a retrieval and are asked to supply feedback regarding the relative importance of the results.

Comparing our approach with these methods of query reformulation, we make a couple of observations. First, query reformulation can be driven by knowledge we have precomputed about a data colection. Given that entities such as papers have abstracts, keywords, and titles, does it make sense to vary the amount of this information in the queries? If we have the equivalent of Table 3 for a collection, we can do a table lookup at run time to determine which formulations are the most promising.

We note that the way we construct queries for the *queryConjunct* method combines aspects of both the Boolean and vector-space models of query formulation into a hybrid approach. In the case of Boolean queries, relevance feedback can lead to new query expressions consisting of term conjuncts such as (Salton, 1997):

$$(Term_i\ AND\ Term_j\ AND\ Term_k)$$

Notice that if we replace any $Term_i$ with $Vector_i$ in the above expression, we have a query expression formulated according to our *queryConjunct* method.

## 7. Conclusions

In this paper, we have shown that we can collect information about reviewers automatically from the Web, and we can use it as a part of a recommendation framework to route papers to reviewers. We treat the problem as one of decomposing reviewer interest and paper contents into information sources, and then of combining the information sources using different query formulations. In our experiments, we compared two ways of formulating queries using content-based information retrieval and one collaborative approach. We have found that the recommendation algorithm using conjunctive queries outperforms the other approaches. We have also looked at using different subsets of the information sources in our algorithms, and in the case of our optimal algorithm, we found that using more information generally lead to better performance.

In a practical setting, the recommendation method of choice is likely to depend on a number of factors ranging from the availability of information to ease of use. On the one hand, our framework provides a more flexible alternative to simple keyword-based searches and a less intrusive alternative to collaborative methods. On the other hand, our methods assume that we can obtain data that are reliable, accurate, and timely. Based on our results, we are optimistic that the Web can provide credible information sources that can be used successfully in the recommendation process.

## 8. Acknowledgments

We extend our thanks to AAAI, the AAAI reviewers, the AAAI paper authors, members of the Rutgers Machine Learning Research Group, and the reviewers of this paper for their inputs in this work.

We note that the following are the property of their respective companies as listed: WHIRL (AT&T Labs – Research), LSI (Telcordia Technologies, Inc.).